\documentclass[journal,twoside,web]{ieeecolor}
\usepackage{tuffc}
\RequirePackage{xcolor}
\usepackage{etoolbox}
\makeatletter
\@ifundefined{color@begingroup}%
  {\let\color@begingroup\relax
   \let\color@endgroup\relax}{}%
\def\fix@ieeecolor@hbox#1{%
  \hbox{\color@begingroup#1\color@endgroup}}
\patchcmd\@makecaption{\hbox}{\fix@ieeecolor@hbox}{}{\FAILED}
\patchcmd\@makecaption{\hbox}{\fix@ieeecolor@hbox}{}{\FAILED}
\usepackage{cite}
\usepackage{amsmath,amssymb,amsfonts}
\usepackage{algorithmic}
\usepackage{graphicx}
\usepackage{textcomp}
\usepackage{wrapfig,colortbl}
\usepackage{xcolor}
\usepackage{soul}
\usepackage{gensymb}
\definecolor{abstractbg}{rgb}{1,0.969,0.914}
\def\BibTeX{{\rm B\kern-.05em{\sc i\kern-.025em b}\kern-.08em
    T\kern-.1667em\lower.7ex\hbox{E}\kern-.125emX}}
\begin{document}
\newcommand{\multrow}[1]{\begin{tabular}{@{}c@{}} #1 \end{tabular}}
\title{Boosting Cardiac Color Doppler Frame Rates with Deep Learning}
\author{Julia Puig, Denis Friboulet, Hang Jung Ling, François Varray, Michael Mougharbel, Jonathan Porée,\\
Jean Provost, Damien Garcia and Fabien Millioz
\thanks{Manuscript submitted for review March 7, 2024.}
\thanks{This work was partially supported by the framework of the LABEX PRIMES (ANR-11-LABX-0063) of Université de Lyon and the LABEX CELYA (ANR-11-LABX-0060) of Université de Lyon, both within the program ”Investissements d’Avenir” (ANR-11-IDEX-0007) operated by the French National Research Agency (ANR).}
\thanks{J. Puig, D. Friboulet, H. J. Ling, F. Varray, D. Garcia and F. Millioz are with Univ Lyon, INSA‐Lyon, Université Claude Bernard Lyon 1, CNRS, Inserm, CREATIS UMR 5220, U1294 69100 Villeurbanne, France (e-mail: julia.puig@creatis.insa-lyon.fr, fabien.millioz@creatis.insa-lyon.fr).}
\thanks{M. Mougharbel, J. Porée and J. Provost are with with the Department of Engineering Physics, Polytechnique Montréal, Montréal, QC H3T 1J4. J. Provost is also with the Montreal Heart Institute, Montréal, QC H1T 1C8, Canada.}
}

\IEEEtitleabstractindextext{%
\fcolorbox{abstractbg}{abstractbg}{%
\begin{minipage}{\textwidth}\rightskip2em\leftskip\rightskip\bigskip
\begin{wrapfigure}[26]{r}{2.7in}%
\hspace{-3pc}\includegraphics[width=2.7in]{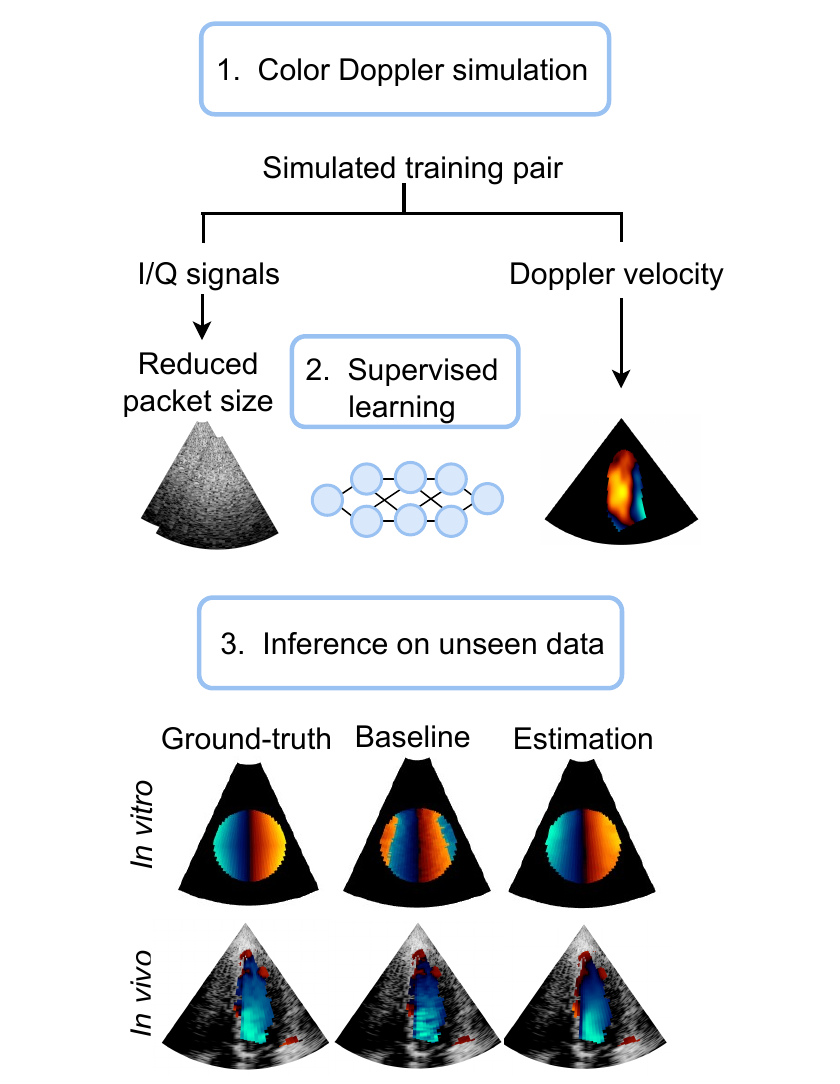}
\end{wrapfigure}%
\begin{abstract}
Color Doppler echocardiography enables visualization of blood flow within the heart. However, the limited frame rate impedes the quantitative assessment of blood velocity throughout the cardiac cycle, thereby compromising a comprehensive analysis of ventricular filling. Concurrently, deep learning is demonstrating promising outcomes in post-processing of echocardiographic data for various applications. This work explores the use of deep learning models for intracardiac Doppler velocity estimation from a reduced number of filtered I/Q signals. We used a supervised learning approach by simulating patient-based cardiac color Doppler acquisitions and proposed data augmentation strategies to enlarge the training dataset. We implemented architectures based on convolutional neural networks. In particular, we focused on comparing the U-Net model and the recent ConvNeXt models, alongside assessing real-valued versus complex-valued representations.  We found that both models outperformed the state-of-the-art autocorrelator method, effectively mitigating aliasing and noise. We did not observe significant differences between the use of real and complex data. Finally, we validated the models on \textit{in vitro} and \textit{in vivo} experiments. All models produced quantitatively comparable results to the baseline and were more robust to noise. ConvNeXt emerged as the sole model to achieve high-quality results on \textit{in vivo} aliased samples. These results demonstrate the interest of supervised deep learning methods for Doppler velocity estimation from a reduced number of acquisitions.
\end{abstract}

\bigskip
\end{minipage}}}

\maketitle

%----------------------------------------------------------------------
% Introduction
%----------------------------------------------------------------------
\section{Introduction}
\label{sec:introduction}
\IEEEPARstart{C}{olor} Doppler echocardiography facilitates the concurrent visualization of cardiac tissue and intracardiac blood movement. In clinical settings, this imaging modality is extensively used as a visual aid to assess cardiac function such as valvular regurgitation, ventricular filling and ventricular diastolic function \cite{mitchell_guidelines_2019}. The frame rate achieved by focused Color Doppler echocardiography is in the range of 10-30 fps assuming an apical long-axis acquisition from the apex to the basis of 12 cm and a number of lines between 20 and 60, depending on the sector width. Considering a standard subject with 80~bpm, this results in 8 to 23 frames per cardiac cycle. However, Krovetz \textit{et al.} \cite{krovetz1974frequency} found that around 25 harmonics are needed to accurately retrieve blood flow accelerations within the left ventricle, indicating that a minimum frame rate of 25 frames per cardiac cycle is necessary to quantitatively assess blood motion during all the phases of the cardiac cycle. Therefore, increasing the temporal resolution of color Doppler could allow a better characterization of intracardiac blood movement and potentially open up new diagnosis possibilities. In particular, Faurie \textit{et al.} \cite{faurie} investigated the use of ultrafast imaging to achieve high-frame rates of around 80~fps for color Doppler imaging. They found that high frame rates allowed to study blood vortex dynamics, and they derived quantitative tools that could be used to assess diastolic impairment. Here, we were interested in achieving high-frame rate color Doppler in the conventional framework of focused imaging.

To generate a color Doppler image, backscattered echoes are acquired in the axial direction (fast-time) at $n$ consecutive instants (slow-time). The radio frequency (RF) signals undergo I/Q demodulation, followed by clutter filtering and beamforming. The number $n$ of slow-time acquisitions is referred to as the packet size. They enable the estimation of the phase shifts along the slow-time axis, from which Doppler velocities are subsequently derived. Phase shifts are usually estimated in two steps \cite{yu_frequency-based_2007}. First, a clutter filter is applied to the I/Q signals to eliminate signals corresponding to tissues and slow-moving artifacts. Then, spatially weighted average autocorrelations are computed to estimate the local phase shifts \cite{loupas}. In practice, the packet size typically ranges around $n=8$. The reliability of the autocorrelation-based estimates significantly relies on the packet size; its reduction compromises robustness to noise and other artifacts. 

In addition to noise, color Doppler can suffer from aliasing, which occurs when velocities exceed the Nyquist limit set by the pulse repetition frequency (PRF). While limited research has been dedicated to dealiasing in echocardiography, it is an essential step in extracting reliable quantitative information from color Doppler. In the context of ultrafast imaging, Posada \textit{et al.} \cite{posada_aliasing_2016} proposed the use of staggered multi-PRF emissions to provide alias-free Doppler velocities. Deliasing has also been addressed as a post-processing problem. Muth \textit{et al.}\cite{muth_aliasing_2011} developed the DeAN algorithm for denoising and dealiasing color Doppler velocity maps. Their method involves unsupervised segmentation of aliased regions and subsequent dealiasing through comparison with neighboring regions. However, the method is prone to failure in cases of deteriorated signal quality.

In recent years, deep learning has led to significant advancements in medical imaging by replacing traditional model-based methods with learning-based approaches. In particular, convolutional neural networks (CNNs) have emerged as pivotal tools in ultrasound imaging, either by introducing entirely learning-based models or by enhancing the performance of existing model-based solutions with a deep learning step \cite{luijten_ultrasound_2023}.

Few works have addressed Doppler velocity estimation with deep learning. For cardiac tissue Doppler estimation, van Sloun \textit{et al.} \cite{vansloun} used an encoder-decoder on the I/Q signals to replace the classical autocorrelator method. Utilizing \textit{in vivo} I/Q data from a porcine model, they derived Doppler estimates that achieved the quality of the autocorrelator method with reduced noise levels. In cardiac color Doppler, Apostolakis \textit{et. al.} \cite{apostolakis_2023} introduced a serial U-Net designed to estimate Doppler velocities from wall-filtered RF signals with reduced ($n=4$) or undersampled packet sizes. They used a training dataset of flow phantom acquisitions whose reference velocities were given by an autocorrelator with a packet size of $n=14$. Lei \textit{et al.} \cite{lei_ultrasonic_2022} proposed a complex-valued CNN dedicated to carotid blood flow velocity estimation from RF signals, which simultaneously addressed clutter removal and velocity estimation, demonstrating promising results on simulated carotid data and \textit{in vivo} experiments. Regarding clutter filtering, Solomon \textit{et al.} \cite{solomon_deep_2020} proposed a deep unfolding algorithm for super-resolution ultrasound applications. They formulated the problem as an iterative Robust Principal Component Analysis (RPCA) and solved it using a deep learning algorithm. In the same context, Brown \textit{et al.} \cite{brown_deep_2020} performed clutter filtering on both \textit{in vitro} and \textit{in vivo} data using a spatio-temporal 3D CNN. 

Regarding dealiasing, in 2020 Nahas \textit{et al.} \cite{nahas_aliasing_2020} trained a U-Net to identify and segment aliased regions in femoral bifurcation color Doppler images using Doppler frequency, power and bandwidth information. Recently, they proposed an accelerated dealiasing technique for high-frame-rate vector Doppler imaging in femoral imaging \cite{nahas_aliasing_2023}. Alternatively, Ling \textit{et al.} \cite{ling_aliasing_2023} compared a deep unfolding method with a data-driven deep learning approach for dealing with aliasing in cardiac color Doppler images. The data-driven approach demonstrated superior results, with both methods outperforming the DeAN method mentioned above. All these dealiasing works can be considered as post-processing methods applied to Doppler velocity maps.

In this study, we propose a deep learning approach as an alternative to the current autocorrelator technique in the context of focused cardiac color Doppler acquisitions. Our goal was to obtain accurate Doppler velocities from clutter-filtered I/Q signals with a reduced packet size of $n=2$. We contended that convolutional neural networks can identify spatial and channel-wise connections that the autocorrelator overlooks, potentially compensating for the packet size reduction. Our main contributions are as follows:
\begin{enumerate}
    \item We generated a training dataset for supervised learning by simulating color Doppler echocardiography with a dedicated pipeline. It included the generation of high-quality Doppler velocity ground truths and the simulation of realistic acquisitions of ultrasound signals. 
    \item We adapted state-of-the-art deep learning architectures for real-time Doppler velocity estimation from I/Q signals with a reduced packet size. We explored the performance of U-Net and ConvNeXt architectures, and the impact of using real-valued versus complex-valued representations.
    \item We introduced a tailored \textit{zooming} data augmentation procedure to increase the number of samples while ensuring signal coherence. In addition, we generated augmented \textit{aliased} samples to make the models inherently robust to aliasing.
    \item We validated the proposed deep learning models using both \textit{in vitro} and \textit{in vivo} sequences. These experiments showed that the deep learning models were able to generalize to domain shift data. They achieved comparable results to the autocorrelator while exhibiting robustness to noise and inherent reduction of aliasing.
\end{enumerate}

%----------------------------------------------------------------------
% Highlights
%----------------------------------------------------------------------
\begin{table*}[!t]
\arrayrulecolor{subsectioncolor}
\setlength{\arrayrulewidth}{1pt}
{\sffamily\bfseries\begin{tabular}{lp{6.75in}}\hline
\rowcolor{abstractbg}\multicolumn{2}{l}{\color{subsectioncolor}{\itshape
Highlights}{\Huge\strut}}\\
\rowcolor{abstractbg}$\bullet$ & We explored the use of deep learning for generating color Doppler images from clutter-filtered echocardiographic I/Q signals. \\
\rowcolor{abstractbg}$\bullet${\large\strut} & We proposed a deep learning model that estimates phase shift from I/Q signals and mitigates aliasing from a reduced packet size. The model was trained solely on \textit{in silico} data and validated on both \textit{in vitro} and \textit{in vivo} sequences. \\
\rowcolor{abstractbg}$\bullet${\large\strut} & It was found that deep learning models can inherently reduce aliasing and are robust to noise induced by a reduced packet size. \\[2em]\hline
\end{tabular}}
\setlength{\arrayrulewidth}{0.4pt}
\arrayrulecolor{black}
\end{table*}

%----------------------------------------------------------------------
% Methods
%----------------------------------------------------------------------
\section{Methods}
\label{methods}
% AC
\subsection{Doppler velocity estimation in color Doppler}
In ultrasound imaging, the received RF signals undergo I/Q-demodulation and beamforming. In the context of Doppler imaging, these resulting complex-valued signals can be organized in arrays of shape $h \times w \times n$, where $h$ and $w$ represent the fast-time and lateral dimensions, respectively, and $n$ denotes the slow-time dimension referred to as the packet size. These processed signals are filtered to remove clutter and retrieve blood-related signals. Blood Doppler velocities are then estimated by calculating the phase shifts of the filtered signals in the slow-time direction, \textit{i.e.} along the third dimension of the array. These phase shifts can be derived by evaluating the argument of the lag-one autocorrelator $(R_1)$ applied to the I/Q signals. We adopted the 2\nobreakdash-D autocorrelator introduced by Loupas \textit{et al.} \cite{loupas} as the baseline method. We smoothed $R_1$ spatially with a Hamming kernel of size $10 \times 4$ pixels to obtain a smoothed autocorrelation $R_1^s$. Finally, the Doppler velocities were computed by converting the arguments to velocities as follows:
\begin{equation*}
    v_D = -\frac{v_N}{\pi} \arg({R^s_1}),
\end{equation*}
where $v_N$ stands for the Nyquist velocity. The reliability of this technique depends on the packet size number, and decreasing it can degrade the estimation quality, especially when dealing with noisy acquisitions.

% Simulations
\subsection{Color Doppler simulations}
\label{simulator}
\begin{figure}[!t]
    \centerline{\includegraphics[width=\columnwidth]{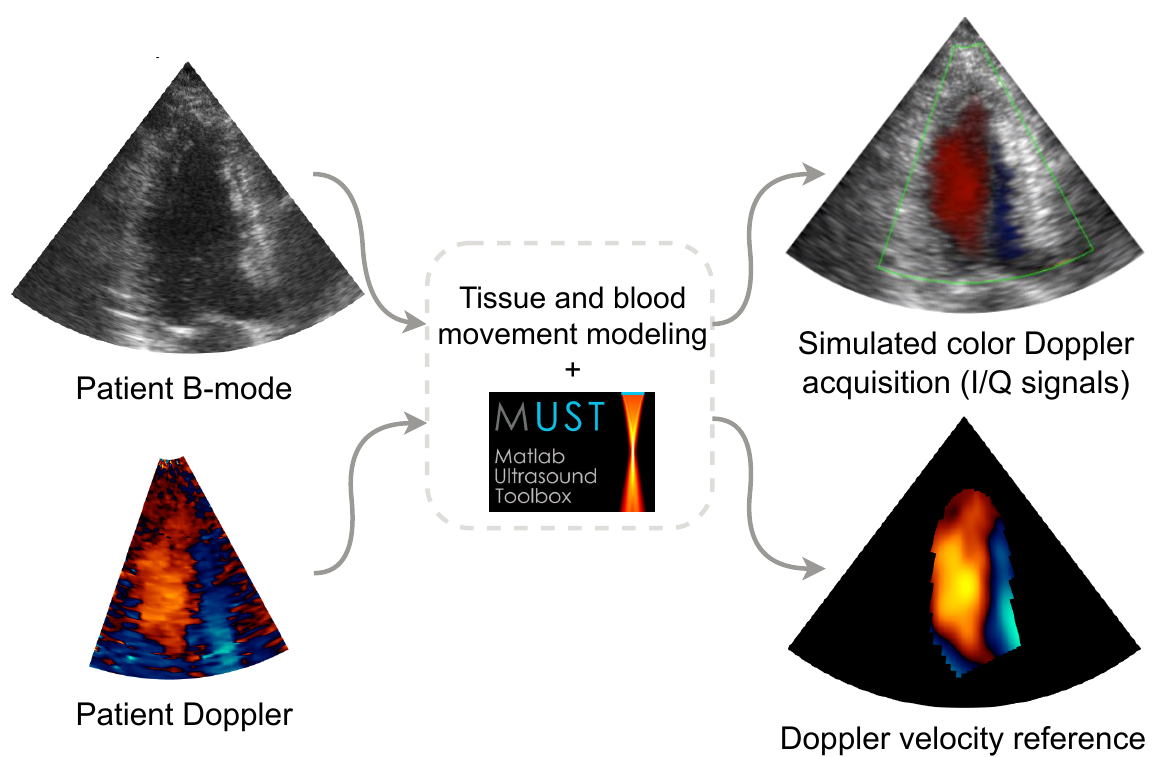}}
    \caption{Inputs and outputs of the color Doppler simulation pipeline.}
    \label{simu_mini_scheme}
\end{figure}

Training supervised deep learning models requires large annotated datasets. In the context of color Doppler, high-quality \textit{in vivo} ground truth is not available. Using directly the patient Doppler data, the models would learn to reproduce the behavior of the standard estimator along with its inaccuracies. Therefore, we decided to rely on a simulation approach that allows to generate a supervised training dataset with reliable ground truths. In this study, we adapted the patient-based color Doppler simulator developed by Sun \textit{et al.} \cite{sun}. Unless otherwise specified, the parameters were consistent with those described in the original paper. A schematic representation  of the inputs and outputs of the simulation pipeline is illustrated in Figure~\ref{simu_mini_scheme}.
Given the patient's B-mode and Doppler information along a cardiac sequence, the simulation pipeline consisted of three main steps.

\subsubsection{Modeling cardiac tissue motion}
For each patient's B-mode image, the inner and outer walls of the left ventricular myocardium were segmented using a deep learning method~\cite{ling_segmentation}. A myocardial mesh was generated from the segmentation masks. Tissue scatterers were randomly distributed in the end-systole frame. Their positions were then propagated frame by frame based on the motion of the myocardial mesh. For additional information, refer to \cite{sun}.

\subsubsection{Modeling blood flow motion}
Patient Doppler velocity frames were treated as independent due to the low frame rate of color Doppler. For each color Doppler frame, a 2\nobreakdash-D intraventricular vector flow map was computed from the patient's Doppler velocities using a physics-based method~\cite{vixege_2021}. The method was used as described in the original paper. Then, for each frame, blood scatterers were distributed in the intraventricular region. Series of $[\text{number of firings} \times n]$ time steps were defined, during which the blood scatterers were displaced according to the 2\nobreakdash-D velocity maps. To enhance realism, fluctuation values were computed from velocity maps and added to velocity values. This process was repeated for all frames of the sequence. From the resulting vector outputs, we used the radial components to build the ground truth Doppler velocity maps $D\in\mathbb{R}^{h \times w}$ in our training dataset.

\subsubsection{Simulation of an ultrasound acquisition}
From the tissue and blood scatterer position maps defined at each time step, scanline-based ultrasound acquisitions were simulated using SIMUS \cite{garcia_simus_2022, cigier_simus_2022} of the Matlab UltraSound Toolbox (MUST) \cite{garcia}. The RF signals were post-processed conventionally with I/Q-demodulation and delay-and-sum beamforming on a $h \times w$ grid. The resulting clutter-free beamformed I/Q signals $S\in\mathbb{C}^{h \times w \times n}$ were the inputs in our training dataset.

% Deep learning
\subsection{Doppler velocity estimation with deep learning}
We explored the use of convolutional neural networks, arguing that the spatial nature of convolutions could compensate for the loss of temporal information and produce better estimates. We focused on two architectures: U-Net and ConvNeXt.

\subsubsection{U-Net}
\begin{figure}[!t]
    \centerline{\includegraphics[width=10cm]{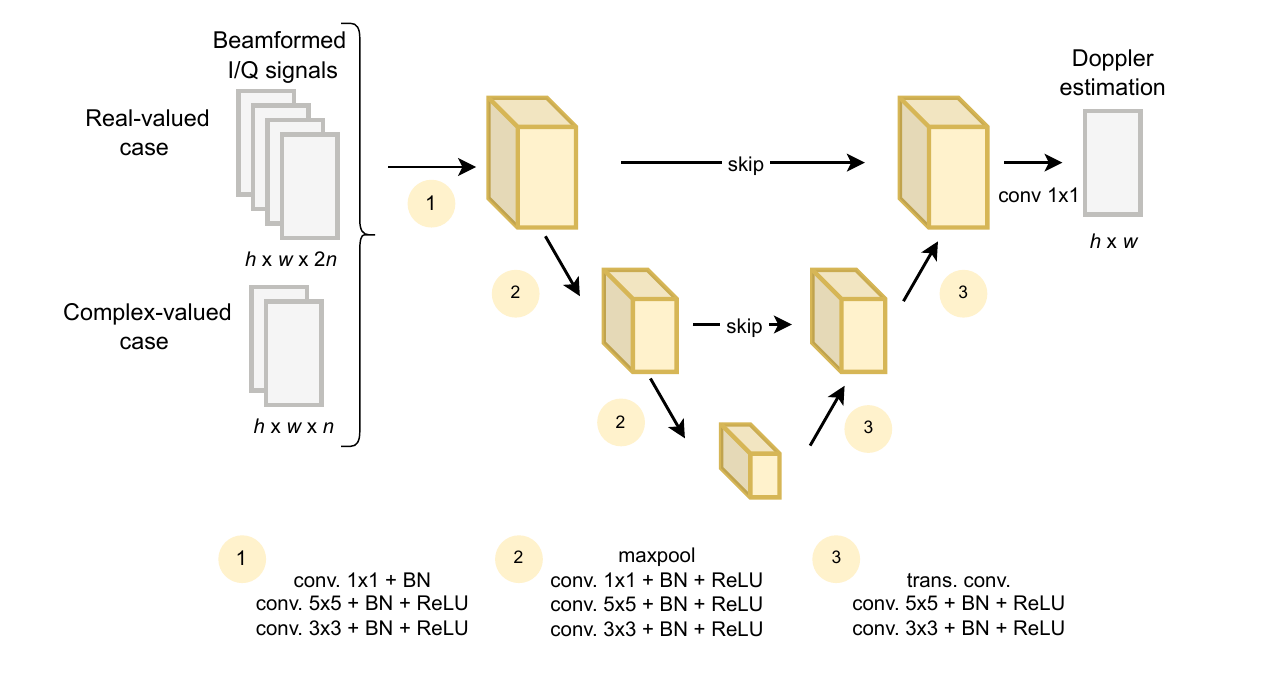}}
    \caption{Architecture of the U-Net-based model.}
    \label{unet}
\end{figure}

The U-Net architecture, introduced in 2015 for medical image segmentation, comprises an encoder-decoder structure with skip connections \cite{unet}. Given that our input $S$ was complex-valued and the output $D$ was real-valued, we examined two architectures: one entirely real-valued and another that was complex-valued in all layers except the last one.
\begin{itemize}
	\item{Real U-Net.}
	The input of the network was in the form $S\in\mathbb{R}^{h \times w \times 2n}$, where the real and imaginary parts of the I/Q signals were concatenated to obtain real representations. We began with the standard U-Net architecture as described in \cite{unet} and made two modifications. Firstly, to ensure a lightweight network, we retained two dimension reduction steps, thereby reducing the network to only 6 layers. Secondly, the kernels of the two convolutions at each U-Net stage were of sizes $5\times5$ and $3\times3$. The number of feature maps was $(32,64,128)$ for the encoder part and symmetrically for the decoder part. A scheme of the architecture is shown in Figure~\ref{unet}. With this configuration, the network had 1.5 million parameters, and the lowest dimension achieved was 45 $\times$ 10.
	\item{Complex U-Net.}
	The input of the network was in the form $S\in\mathbb{C}^{h \times w \times n}$. The network features mirrored those of the Real U-Net, with the following distinctions. We employed complex-valued convolutions using the PyTorch framework. In particular, the network weights were complex-valued in all layers except the last one, where the real and imaginary parts of the feature map were concatenated before applying the real-valued weights. For the complex-valued activation function, we used the Complex ReLU, as advised by Trabelsi \textit{et al.} \cite{trabelsi_deep_2018}. Finally, we used the 2\nobreakdash-D complex batch normalization available in \cite{matthes_complexbatchnorm_2021}, implemented following the recommendations of \cite{trabelsi_deep_2018} in order to obtain equal variance in both the real and imaginary parts. The network contained the same number of parameters as the Real U-Net (1.5 M), but the convolutional operations were four times more computationally intensive.
\end{itemize}

\subsubsection{ConvNeXt}
\begin{figure*}[!t]
    \centerline{\includegraphics[width=18cm]{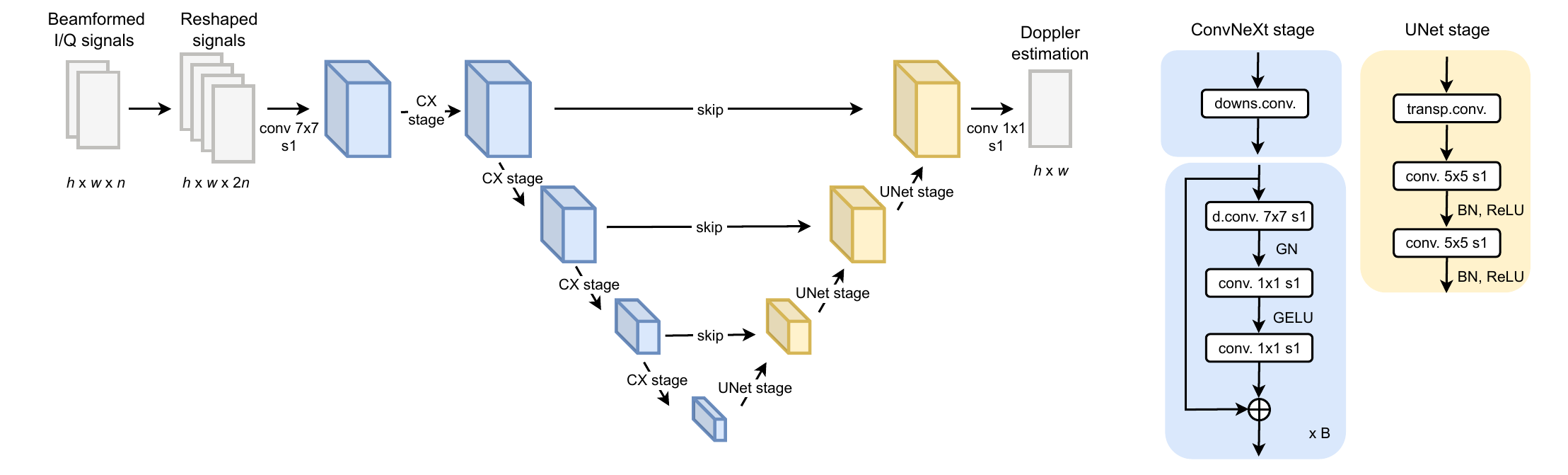}}
    \caption{Architecture of the ConvNeXt-based model. Initially, I/Q signals were reshaped by concatenating their real and imaginary parts. A first set of $7 \times 7$ convolutions was applied. Following this, four ConvNeXt stages were implemented, each containing $(3,3,9,3)$ blocks, with a down-sampling step in all stages except the first. The number of feature maps was $(32,64,128,256)$ for the encoder part and symmetrically for the decoder part. "d.conv.", "trans.conv." and "downs.conv." stand for depthwise, transposed and pooling convolution, respectively. "s" indicates the convolution stride.}
    \label{convnext}
\end{figure*}

We then considered ConvNeXt \cite{liu2022convnet}, a recent convolutional neural network architecture. Since 2020, hierarchical vision Transformer architectures have demonstrated superior performance over convolutional models in various computer vision tasks. However, Transformers models suffer from being data-intensive. As a solution, ConvNeXt was introduced in 2022. It is a convolutional-based model that integrates strategies from Transformer models, resulting in an architecture that outperforms state-of-the-art Transformers while remaining data efficient. One of its key features is the separate handling of spatial and channel information, akin to the self-attention mechanism in Transformers. This property is achieved through the use of depthwise separable convolutions. We adapted the ConvNeXt architecture to our problem by utilizing a U-Net architecture with ConvNeXt as its backbone (\textit{i.e.}, ConvNeXt layers served as feature extractors in the encoder part). A schematic representation of the considered architecture is depicted in Figure~\ref{convnext}.
 With this configuration, the network comprised 4.7 million parameters, and the lowest dimension attained was 11 $\times$ 5.

% Data augmentation
\subsection{Data augmentation}
In computer vision, a common technique for data augmentation involves applying simple transformations to available images to increase the size and variability of the training set, resulting in a more robust model. However, for ultrasound data, the options for realistic data augmentation are limited due to the specific geometry of the acquisition process. Additionally, some augmentation techniques require interpolation, which may significantly alter the signal information. Therefore, we only used vertical flipping and a custom zooming procedure.

\subsubsection{Zoomed samples}
\label{zoomed_samples}
\begin{figure}[!t]
    \centerline{\includegraphics[width=10cm]{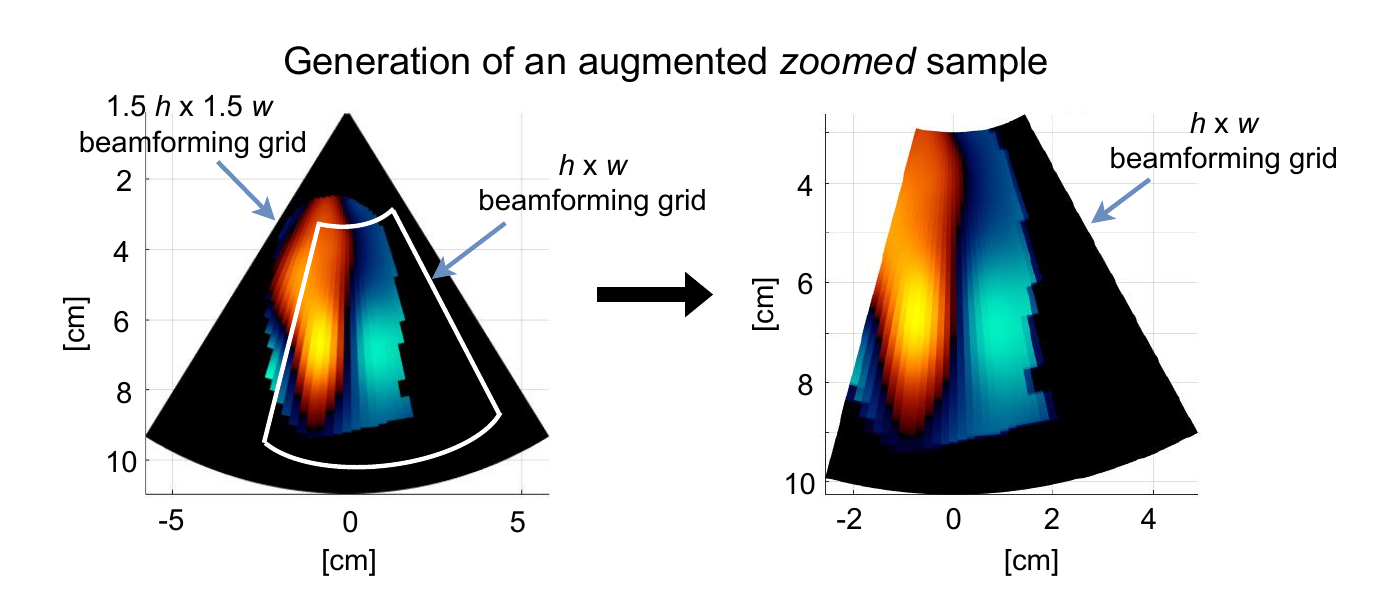}}
    \caption{Illustration of the zooming data augmentation procedure.}
    \label{zooming_procedure}
\end{figure}

Standard simulations had $h \times w$ shape, where $h$ and $w$ refer to the original number of fast-time points and number of firings, respectively. Zooming was achieved by simulating color Doppler acquisitions with a finer grid than the one used for standard simulations, and then cropping the result to an $h \times w$ shape at a random location. This was done by repeating simulations and increasing the number of fast-time points with a 1.5 ratio and the number of focused firings with a 1.5 ratio. The procedure is illustrated in Figure~\ref{zooming_procedure}.

\subsubsection{Aliased samples}
\label{aliased_samples}
Aliasing patterns appear in Doppler velocity maps when the maximum speed exceeds the Nyquist velocity. In such cases, phase unwrapping methods can be applied to the Doppler velocity maps as a post-processing step, as described in the introduction. In contrast to these post-processing approaches, we aimed to make the Doppler estimation inherently robust to aliasing. This was achieved by providing the deep learning models with new training pairs $(S,D)$, where $S$ represents the I/Q aliased signal and $D$ denotes the corresponding alias-free Doppler ground truth. The aliased samples were obtained during simulations by randomly reducing the original PRF values of each sequence by a factor in the $[0.4, 0.6]$ range. Each simulated aliased sample was considered as an \textit{augmented} sample and added to the dataset.

%----------------------------------------------------------------------
% Experiments
%----------------------------------------------------------------------
\section{Experiments}
\label{experiments}
% Data acquisition
\subsection{Data acquisition}
% Simulations experiment
\subsubsection{\textit{In silico} experiment}
Our simulations, described in Section~\ref{simulator}, aimed to neglect the influence of clutter through the exclusive modeling of tissue-free scenarios. As input, we used patient color Doppler echocardiographic data acquired in a previous study \cite{mehregan_doppler_2014}. Examinations adhered standard echocardiographic procedures, resulting in a diverse patient population encompassing individuals with and without cardiac disease. Due to randomization and anonymization, demographic and clinical details were unknown during the present work. The sequences, captured from the apical three-chamber view, incorporated both B-mode and Doppler velocity information. We obtained cardiac sequences of 37 patients, each containing at least one complete cardiac cycle (total: 2,576 samples). From these patient echocardiographic data, we simulated 2,576 training pairs. The probe and acquisition parameters are summarized in Table~\ref{parameters}.
\begin{table}
    \centering
    \caption{Parameters of the simulated ultrasound acquisition}
    \label{table}
    \setlength{\tabcolsep}{3pt}
    \begin{tabular}{ll}
    \hline\hline
    Parameter & 
    Value \\
    \hline
    simulated probe& Verasonics P4-2v \\
    central frequency ($f_c$) & $2.7 \times 10^6$ Hz \\
    pitch & $300 \times 10^{-6}$ m \\
    \#elements & 64 \\
    bandwidth & 74\% \\
    \#cycles B-mode & 1 \\
    \#cycles Doppler & 6 \\
    PRF & 4500 - 7700 Hz \\
    sector width & 60$\degree$ - 90$\degree$ \\
    depth & 10 - 17 cm \\
    sampling frequency & 4$f_c$ \\
    acquisition type & focused waves \\
    \#firings B-mode & 120 \\
    \#firings Doppler & 40 \\
    packet size ($n$) & 8 \\
    \hline\hline
    \end{tabular}
    \label{parameters}
\end{table}
During the simulations, we leveraged patient-specific parameters as pulse repetition frequency (PRF), sector angle, and depth. The simulated RF signals were demodulated into I/Q signals and beamformed into a $180 \times 40$ grid using delay-and-sum. Each training pair $(S,D)$ consisted of I/Q signals $S\in\mathbb{C}^{180 \times 40 \times n}$ and Doppler velocity references $D\in\mathbb{R}^{180 \times 40}$. 

\begin{figure}[!t]
    \centerline{\includegraphics[width=\columnwidth]{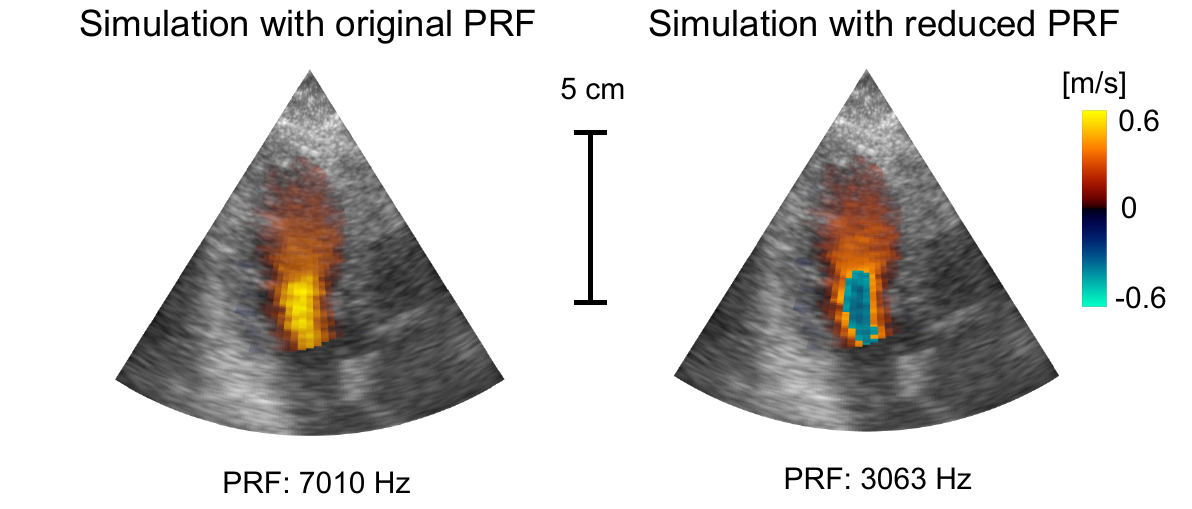}}
    \caption{Example of a simulated echocardiography acquisition with its original and decreased PRF. The Doppler velocities were estimated using an autocorrelator in both cases.}
    \label{simulation_example}
\end{figure}

The simulated dataset is referred to as Original-Set and described in Table~\ref{simulated_datasets}.
\begin{table}[t]
    \small
    \centering
    \caption{Description of the three simulated datasets.}
    \label{simulated_datasets}
    \begin{tabular}{cccc}
    \hline \hline
        & Total & Non-aliased & Aliased \\ \hline
        Original-Set & 2,576 & 2,552 & 24 \\
        Aliased-Set & 1,142 & 0 & 1,142 \\
        Zoomed-Set & 2,576 & 1,883 & 693 \\\hline \hline
    \end{tabular}
\end{table}
A sample form Original-Set is shown in Figure~\ref{simulation_example}.
Only 24 samples contained aliased regions, accounting for less than 0.001\% of the total dataset pixels. This was circumvented with the simulation procedure described in Section~\ref{aliased_samples} to increase the number of aliased samples by artificially decreasing the PRF. The set of aliased samples obtained is referred to as Aliased-Set in the sequel and is described in Table~\ref{simulated_datasets}. An example of a sample simulated with both its original and decreased PRF is depicted in Figure~\ref{simulation_example}. Finally, a third dataset was simulated to provide for augmented \textit{zoomed} samples as described in Section~\ref{zoomed_samples}. This dataset is referred to as Zoomed-Set and is described in Table~\ref{simulated_datasets}.

From these three simulated datasets, we defined two training datasets: Train-Set-1, including Original-Set and non-aliased augmented samples of Zoomed-Set; and Train-Set-2, comprising all simulated datasets. They are described in Table~\ref{training_datasets}.
\begin{table}[t]
    \small
    \centering
    \caption{Description of the two training datasets.}
    \label{training_datasets}
    \begin{tabular}{ccccc}
    \hline \hline
        & Source data & Total & Non-aliased & Aliased\\ \hline
        Train-Set-1 & \multrow{Original-Set \\Zoomed-Set} & 4,459 & 4,435 & 24 \\ \hline
        Train-Set-2 & \multrow{Original-Set \\Aliased-Set \\Zoomed-Set} & 6,294 & 4,435 & 1,859 \\\hline \hline
    \end{tabular}
\end{table}

Finally, samples with reduced packet size were created by selecting two consecutive frames in the slow-time.

% Phantom experiment
\subsubsection{\textit{In vitro} experiment}
We designed a rotating disk experiment to measure Doppler velocities from a color Doppler acquisition. We constructed a 6 cm diameter agar disk and connected it to a rotating motor. A Verasonics P4-2v cardiac probe was positioned at 8 cm from the center of the disk. To minimize clutter and artifacts, we submerged the entire setup in a water tank.

The acquisition parameters were set as described in Table~\ref{parameters}, with the sector width, depth and PRF configured to $50^\circ$, 12~cm and 6,000~Hz, respectively. The packet size was $n=32$. We ran a total of 12 experiments, varying the maximum outer speed between 0.1 and 1.13 m/s, mirroring intracardiac blood velocities. The Nyquist velocity was $v_N=0.85$ m/s for all experiments, resulting in four experiments displaying aliased regions.

Samples with reduced packet size were created by selecting two consecutive frames in the slow-time. The ground truth Doppler velocity maps used to compute the models' estimation errors during inference were obtained from the known rotation speed of the phantom.

% Volunteer experiment
\subsubsection{\textit{In vivo} experiment}
Cardiac echo-Doppler focused acquisitions were conducted on six healthy volunteers. The study received approval from the ethics and research committee of Polytechnique Montréal (CER\nobreakdash-2122\nobreakdash-54\nobreakdash-D). In accordance with the simulation and phantom experiments, the P4-2v Verasonics probe was used with identical parameters as those used for the phantom. Data were acquired in the apical three-chamber view, including both B-mode and Doppler information. The packet size was set to $n=16$. The ultrasound signals were post-processed and beamformed on a $180 \times 40$ grid. We obtained 8 to 16 color Doppler image samples from each volunteer.

To be consistent with the clutter-free setting of both \textit{in silico} and \textit{in vitro} experiments, we applied an SVD-based clutter filter.  We discarded the first four largest singular values to isolate the blood I/Q signals. We segmented the left ventricular endocardium to mask off I/Q signals not pertaining to intraventricular blood. Samples with low Doppler power in over 70\% of the ventricular region were discarded, resulting in 50 selected samples. Only one of the selected samples contained aliased regions. To assess the method's robustness, we deliberately introduced aliasing by considering one out of two (or three) frames in the slow-time, thereby artificially decreasing the PRF by a factor of two (or three). We applied this procedure to the 50 selected samples, providing 20 new samples with several aliased regions.

Samples with reduced packet size were generated by selecting two consecutive frames in the slow-time. In the \textit{in vivo} experiments, ground truth Doppler velocity maps were not available. Hence, Doppler velocity maps obtained with the autocorrelator using a packet size of $n=8$ were used as references.

% Deep learning implementation
\subsection{Deep learning implementation details}
% Training settings
\subsubsection{Training settings}
\label{training_settings}
Because the ultrasound data were simulated with patient-specific PRFs, the Nyquist velocity varied from sample to sample. Therefore, we opted to use phase shifts as the ground truth instead of Doppler velocities to ensure standardized inputs across samples. Each sample was normalized with respect to its maximum complex modulus across the slow-time. For greater robustness, we trained the network using 9-fold cross-validation, where samples from a given patient were assigned to the same fold. Once the test fold was chosen, the split of samples in training/validation was set to 9/1 while ensuring that samples from a given patient remained in the same split. The network weights were initialized with the Xavier distribution. The loss function was the masked mean squared error, where the mask corresponded to the left ventricular cavity. The optimizer was AdamW, and the batch size was set to 16. The initial value of the learning rate was set to $10^{-3}$ and decreased following a plateau scheduler with a patience of 10 epochs. Trainings were performed using the PyTorch library on an NVIDIA V100 GPU with 16 GB of memory.
% Inference
\subsubsection{Inference}
Inference on the \textit{in silico} dataset was executed using the 9-fold cross-validation procedure, thus obtaining a prediction for each sample of the dataset. Ensemble inference was carried out for both the \textit{in vitro} and \textit{in vivo} datasets, by calculating the median value predicted for each pixel by the nine models of the cross-validation.
% Metrics
\subsection{Metrics}
To gauge the accuracy of Doppler velocity estimation, we calculated the root mean squared error (RMSE) between the ground truth and predicted maps, and examined its mean and standard deviation. Note that for the \textit{in vivo} experiments, we refer to this metric as the root mean squared difference (RMSD) as we only had access to an imperfect reference, not the ground truth. 

%----------------------------------------------------------------------
% Results
%----------------------------------------------------------------------
\section{Results}
\label{results}
% In silico
\subsection{\textit{In silico} results}
The performance of our methods in Doppler velocity estimation, when trained on Train-Set-1 and using a packet size $n=2$, is presented in Table~\ref{insilico_results_1}. For all methods, the metrics were computed exclusively within the left ventricle.
\begin{table}[t]
    \small
    \centering
    \caption{Doppler velocity estimation results of the baseline method and the three proposed deep learning methods, trained on Train-Set-1 and evaluated on Original-Set, Aliased-Set and aliased pixels of Aliased-Set. The results correspond to the mean ± standard deviation of the RMSE [cm/s].}
    \label{insilico_results_1}
    \begin{tabular}{cccc}
    \hline \hline
        Method & Original-Set & Aliased-Set & Aliased pixels \\ \hline
        Autocorrelator & $3.7 \pm 2.6$ & $16 \pm 9.8$ & $57 \pm 23$ \\
        Real U-Net & $1.8 \pm 0.9$ & $5.4 \pm 5.3$ & $14 \pm 14$ \\
        Complex U-Net & $1.8 \pm 0.8$ & $\pmb{5.3 \pm 5.1}$ & $\pmb{14 \pm 13}$ \\
        ConvNeXt & $\pmb{1.7 \pm 1.1}$ & $6.1 \pm 6.3$ & $16 \pm 15$ \\\hline \hline
    \end{tabular}
\end{table}

All deep learning approaches outperformed the autocorrelator. Real U-Net and Complex U-Net provided comparable results, with ConvNeXt slightly outperforming both. In addition, the variability of the three models was significantly lower than that of the autocorrelator.

We examined the samples with higher errors for both the autocorrelator and the deep learning models. For the former, large estimation errors mainly occurred in two situations: noisy signals and in the presence of aliasing. For the deep learning models, large errors were observed in two specific settings: in a few pixels near the valves and the presence of aliasing across a large region of the ventricle. To mitigate the effects of aliasing, we trained our models on Train-Set-2. The results of the estimation are presented in Table~\ref{insilico_results_2}.
\begin{table}[t]
    \small
    \centering
    \caption{Doppler velocity estimation results of the baseline method and the three proposed deep learning methods trained on Train-Set-2 and evaluated on Original-Set, Aliased-Set and aliased pixels of Aliased-Set. The results represent the mean ± standard deviation of the RMSE [cm/s].}
    \label{insilico_results_2}
    \begin{tabular}{cccc}
    \hline \hline
        Method & Original-Set & Aliased-Set & Aliased pixels \\ \hline
        Autocorrelator & $3.7 \pm 2.6$ & $16 \pm 9.8$ & $57 \pm 23$ \\
        Real U-Net & $2.0 \pm 0.8$ & $2.4 \pm 1.1$ & $3.7 \pm 3.1$ \\
        Complex U-Net & $2.0 \pm 0.8$ & $2.5 \pm 1.3$ & $3.7 \pm 3.0$ \\
        ConvNeXt & $\pmb{1.8 \pm 0.7}$ & $\pmb{2.1 \pm 1.0}$ & $\pmb{3.3 \pm 2.4}$ \\\hline \hline
    \end{tabular}
\end{table}

We noted that when testing on Original-Set, we maintained a consistent performance compared to the models trained on Train-Set-1. Moreover, when testing on Aliased-Set, we observed a substantial reduction in error compared to the autocorrelator method, indicating that the models are capable to alleviate aliasing to some extent.

\begin{figure*}[!t]
    \centerline{\includegraphics[width=16cm]{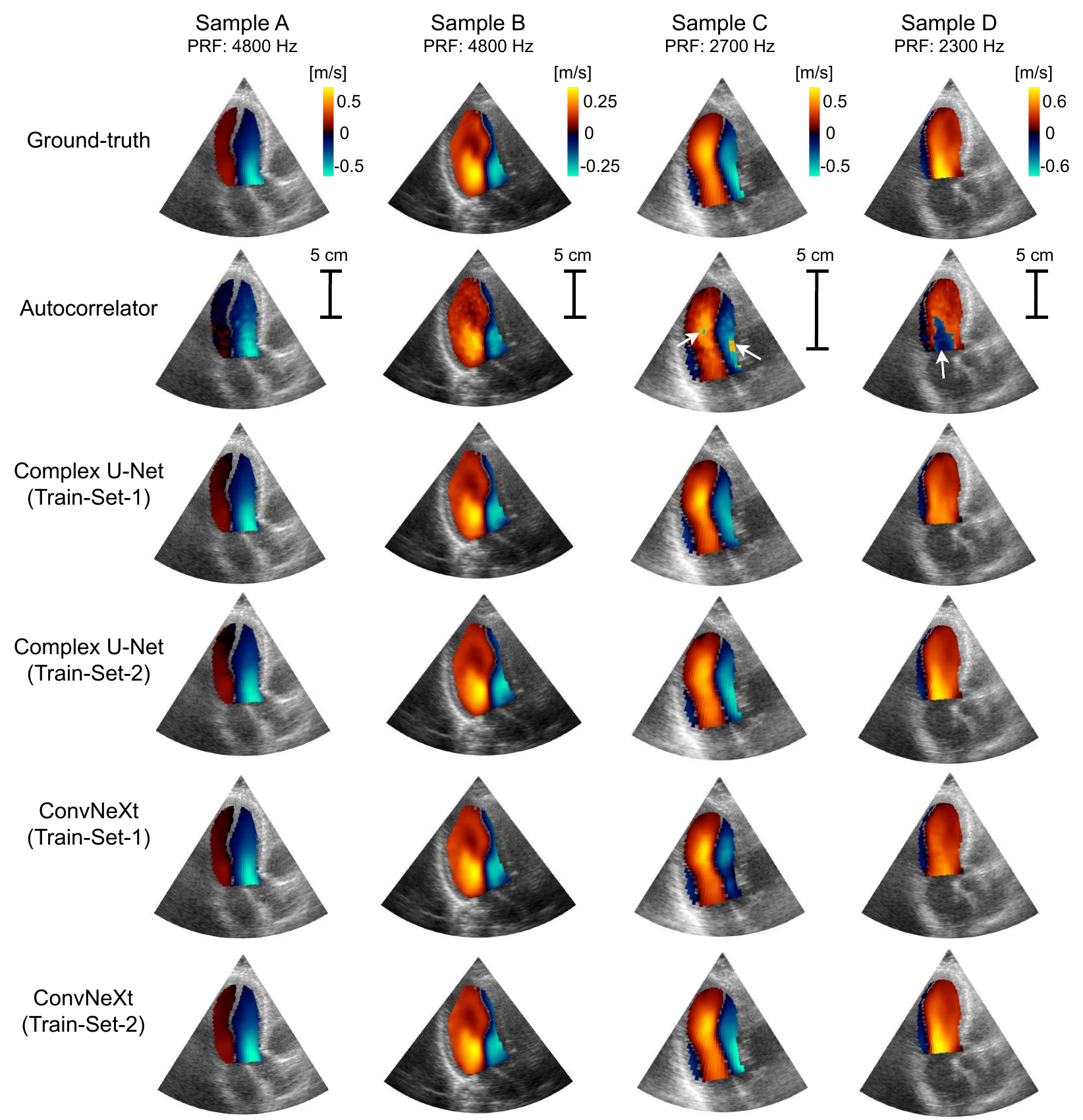}}
    \caption{Doppler velocity estimations for four \textit{in silico} samples. B-mode and color Doppler images were simulated from the data of four different patients. Samples A and B belonged to Original-Set and had velocities below the Nyquist limit. Samples C and D had velocities belonged to Aliased-Set above the Nyquist limit. Aliased regions are indicated with white arrows.}
    \label{insilico_examples}
\end{figure*}

Illustrative examples are shown in Figure~\ref{insilico_examples}.
We report predictions obtained with ConvNeXt (the best-performing model) and Complex U-Net. The Real U-Net provided results similar to the Complex U-Net. As explained above, our models trained with both Train-Set-1 and Train-Set-2 achieved accurate estimates in the absence of aliasing (Sample A and Sample B). In such cases, we observed that the autocorrelator was more susceptible to noise. When encountering small aliased regions (Sample C), the models trained with Train-Set-1 either effectively removed aliasing (see the Complex U-Net estimation), or tended to smooth the region, leading to velocity underestimation (see the right part of the ConvNeXt estimation). When trained with Train-Set-2, the estimation was closer to the ground truth (see the ConvNeXt estimation). In the case of larger aliased regions as in Sample D, the models trained on Train-Set-1 systematically smoothed the aliased region and hence underestimated velocities, whereas when trained with Train-Set-2 they yielded results closely aligned with the ground truth.

% In vitro
\subsection{\textit{In vitro} validation}
\begin{figure}[!t]
    \centerline{\includegraphics[width=7cm]{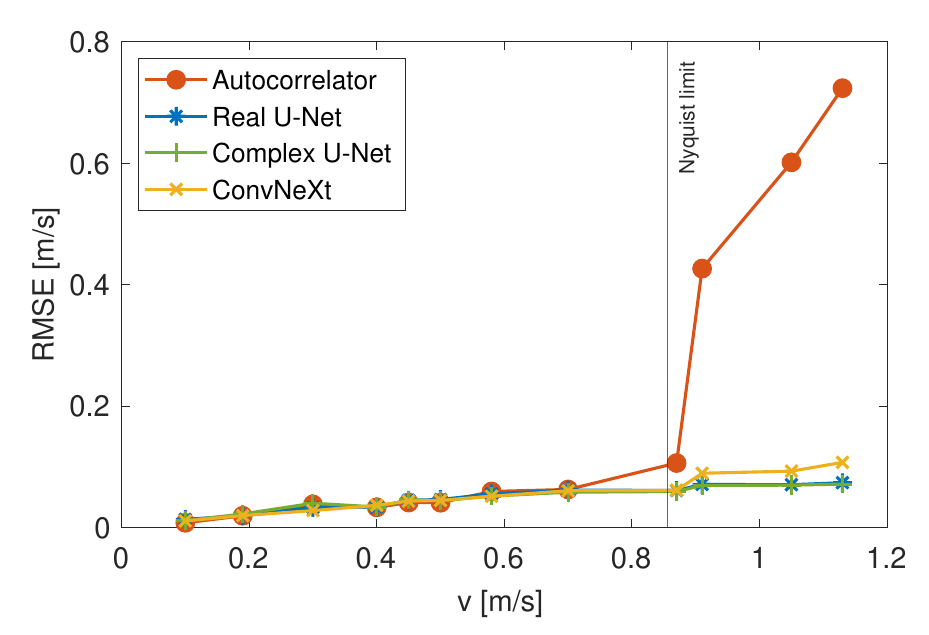}}
    \caption{Evolution of the RMSE with respect to the outermost speed of the phantom, obtained by the autocorrelator and deep learning methods. The RMSE is computed in the phantom area only. The Nyquist limit is indicated with a gray line.}
    \label{invitro_results}
\end{figure}

Figure~\ref{invitro_results} shows the RMSE obtained with the baseline and deep learning-based methods for all 12 \textit{in vitro} experiments, using the same packet size $n=2$. All models were trained on Train-Set-2. The RMSE is depicted as a function of the phantom outermost speeds. The Nyquist velocity $v_N$ was kept unchanged across all experiments. We observed that our methods gave state-of-the-art results for $v<v_N$ and showed improved performance when $v>v_N$. Real U-Net and Complex U-Net gave comparable results, while ConvNeXt slightly underperformed them in terms of aliasing reduction for high rotation speeds.

\begin{figure}[!t]
    \centerline{\includegraphics[width=9cm]{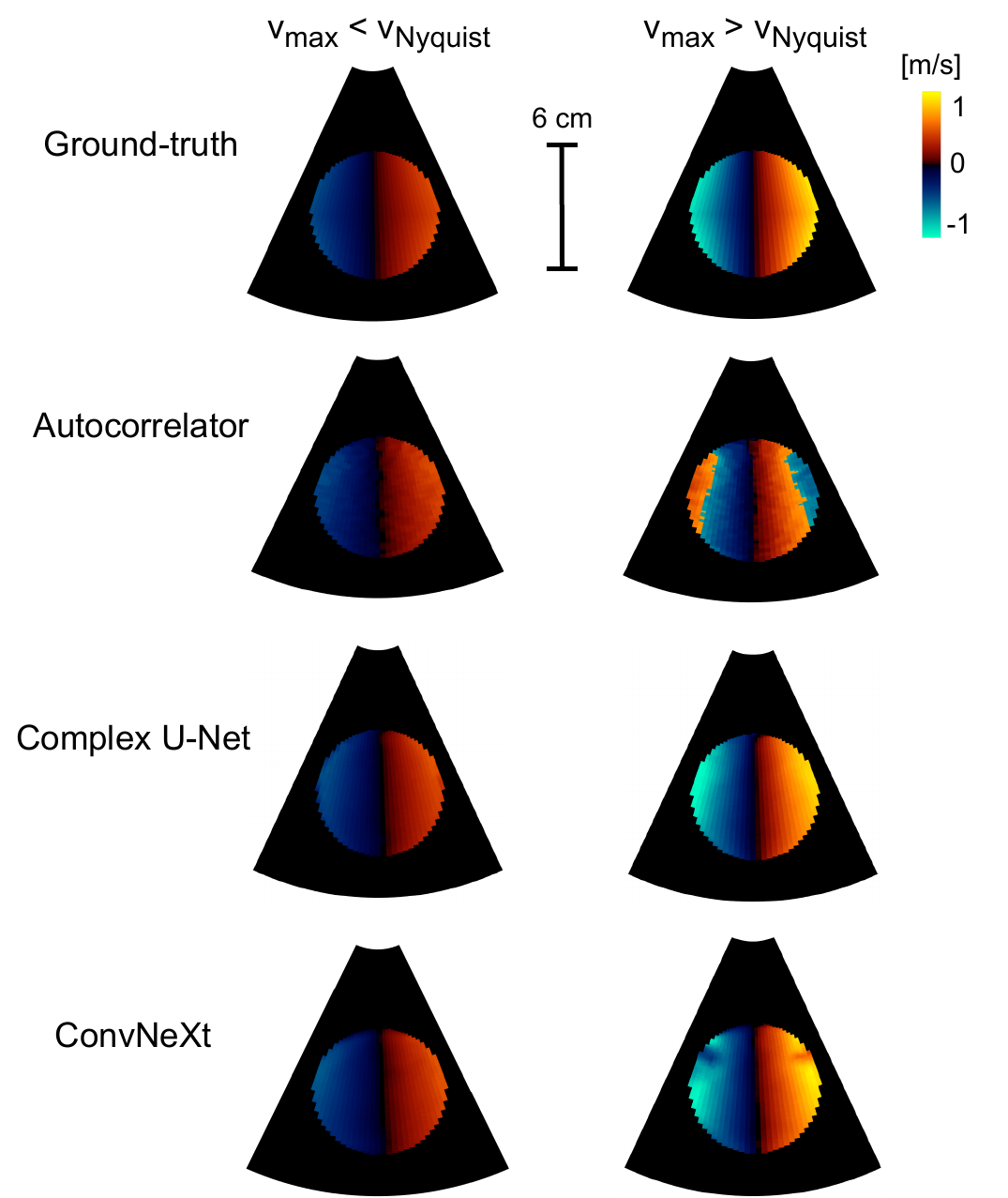}}
    \caption{Doppler velocity estimation results for two \textit{in vitro} experiments with different rotating velocity. In the first one, $v_{\text{max}}$ = 0.45 m/s. In the second one, $v_{\text{max}}$ = 1.15 m/s. The Doppler fields are masked to show the phantom region only.}
    \label{disk_complex}
\end{figure}

Estimation maps for two of the experiments are presented in Figure~\ref{disk_complex}.
In the first experiment,  where the maximum velocity was below the Nyquist limit, we observed estimations closely aligned with the ground truth for all our networks. Moreover, our estimates were less noisy compared to those of the autocorrelator. In the second experiment, where the maximum velocity exceeded the Nyquist limit, the autocorrelator estimation clearly revealed the aliased regions. In contrast, our methods significantly reduced the aliased patterns. However, ConvNeXt displayed small noisy artifacts in the aliased regions.

% In vivo
\subsection{\textit{In vivo} validation}
Estimation results on the \textit{in vivo} sequences are reported in Table~\ref{invivo_results}. We summarized results on 50 \textit{standard} volunteer samples, and on 20 aliased samples, and the reference used for all RMSD calculations was the autocorrelation result with $n=8$. All models were trained on Train-Set-2.
\begin{table}[t]
    \small
    \centering
    \caption{Doppler velocity estimation results of the baseline method and the three proposed deep learning methods, evaluated on standard volunteer and aliased volunteer samples. The results represent the mean ± standard deviation of the RMSD [cm/s].}
    \label{invivo_results}
    \begin{tabular}{ccc}
        \hline \hline
        Method & \textit{Standard} samples & Aliased samples \\ \hline
        Autocorrelator & $\pmb{5.5 \pm 4.5}$ & $17 \pm 8.8$ \\
        Real U-Net & $7.7 \pm 5.4$ & $9.8 \pm 4.3$ \\
        Complex U-Net & $7.3 \pm 4.9$ & $11 \pm 4.4$ \\
        ConvNeXt & $7.3 \pm 4.6$ & $\pmb{6.9 \pm 3.6}$  \\\hline \hline
    \end{tabular}
\end{table}

\begin{figure*}[!t]
    \centerline{\includegraphics[width=17cm]{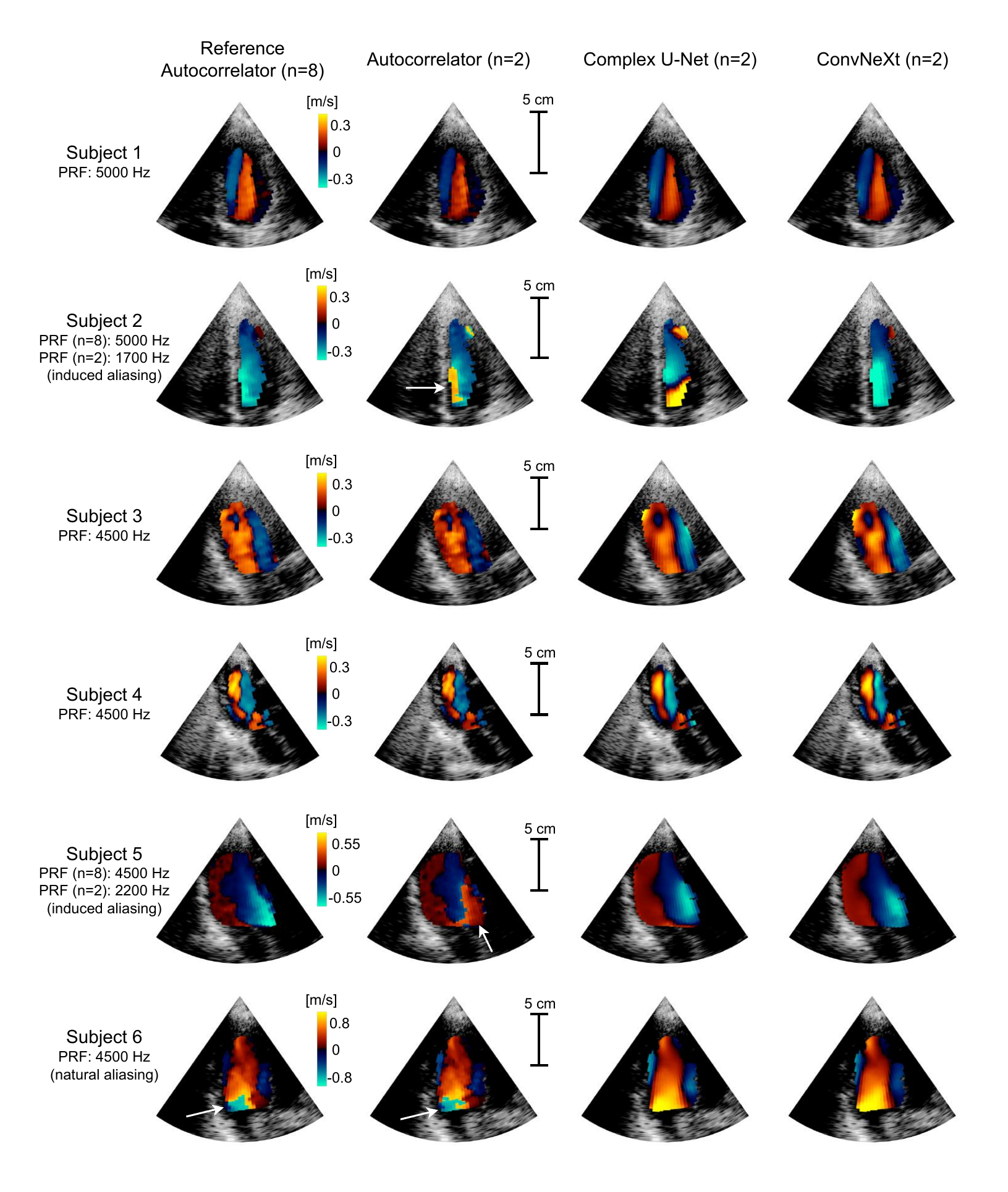}}
    \caption{Doppler velocity estimation results on a \textit{in vivo} sample of six different subjects, using the Autocorrelator with a packet size of 8 as reference. In Subject 2 and Subject 5, samples with $n=2$ were constructed with a decreased PRF to induce aliasing. The sample from Subject 6 naturally contained aliasing. Aliased regions are indicated with white arrows.}
    \label{invivo_predictions}
\end{figure*}

For \textit{standard} samples, the autocorrelator with $n=2$ produced the results closest to the reference. Figure~\ref{invivo_predictions} shows illustrative examples for Complex U-Net and ConvNeXt on one sample of each of the six subjects. From the samples of Subject 1 and Subject 4, all three methods displayed a similar estimation than the reference but we observed that the two deep learning methods behaved as denoisers, unlike the autocorrelator with $n=2$. Moreover, in the sample from Subject 3, the two deep learning estimations were also smoother than the reference obtained with $n=8$.

For aliased samples, as anticipated, the autocorrelator exhibited a substantial deviation from the reference due to aliasing, as reported in Table~\ref{invivo_results}. Among the deep learning methods, ConvNeXt gave results significantly closer to the reference compared to the Real U-Net and the Complex U-Net. In Figure~\ref{invivo_predictions}, the sample from Subject 2 exemplifies a scenario with induced aliasing. The ConvNeXt reduced aliasing and produced an overall estimation very close to the reference, while the Complex U-Net yielded relatively large errors in the apical region and near the valve. In the sample from Subject 5, both Complex U-Net and ConvNeXt successfully reduced the induced aliasing. Finally, the sample from Subject 6 displays a sample with real aliasing during filling. Both the Complex U-Net and the ConvNeXt successfully reduced aliasing.

% Computational complexity and speed
\subsection{Computational time}
Table~\ref{complexity} reports the sizes of the models and their training and inference times. With the same parameter configuration, Complex U-Net required significantly more training and inference time than Real U-Net. In contrast, ConvNeXt offered the most advantageous trade-off between the number of parameters and the training and inference times. The inference times of the three models were sufficiently low to support real-time imaging.
\begin{table}[t]
    \small
    \centering
    \caption{Comparison of the complexity of the autocorrelator and the three deep learning models. The training times are reported for the training of one fold, assuming that all folds can be trained in parallel. The inference times are reported for the inference of in vivo frames. Training and inference were both conducted on GPU for deep learning models and on CPU for the autocorrelator.}
    \label{complexity}
    \begin{tabular}{{cccc}}
    \hline \hline
        Model & Parameters & Training time & Inference time \\ \hline
        Autocorrelator & x & x & $\sim$ 1.6 ms \\
        Real U-Net & 1.5 M & $\sim$ 0.5 h & $\sim$ 2.5 ms \\
        Complex U-Net & 1.5 M & $\sim$ 5 h & $\sim$ 18 ms \\
        ConvNeXt & 4.7 M & $\sim$ 1 h & $\sim$ 5.3 ms \\
    \hline \hline
    \end{tabular}
\end{table}

%----------------------------------------------------------------------
% Discussion
%----------------------------------------------------------------------
\section{Discussion}
\label{discussion}
% commenting the results
\subsection{Doppler velocity estimation with deep learning}
We achieved overall reliable real-time Doppler velocity estimates from a reduced packet size $n=2$ with all proposed deep learning models. This held true for alias-free samples across \textit{in silico}, \textit{in vitro} and \textit{in vivo} data (Table~\ref{insilico_results_1}, Figure~\ref{invitro_results} and Table~\ref{invivo_results}). When working with aliased sequences, we noticed that the models trained on alias-free data (Train-Set-1) tended to smooth the aliased regions (Sample D of Figure~\ref{insilico_examples}), rendering post-processing dealiasing methods less suitable. Augmenting the training dataset with aliased simulations notably improved the model's ability to mitigate aliasing in both \textit{in silico} and \textit{in vitro} data (Sample D of Figure~\ref{insilico_examples} and the second experiment of Figure~\ref{disk_complex}). For \textit{in vivo} data, only ConvNeXt reliably reduced aliasing in all samples (Table~\ref{invivo_results} and Figure~\ref{invivo_predictions}).

Regarding the autocorrelation method, our analysis revealed that our deep learning models yielded comparable quantitative results on \textit{in vitro} and \textit{in vivo} validations (Figure~\ref{invitro_results} and Table~\ref{invivo_results}). We also observed that this method remained stable when the packet size was reduced to $n=2$, albeit producing some level of noise on the velocity maps (Subjects 1 and 4 in Figure~\ref{invivo_predictions}). However, in some cases, the estimation displayed a noticeable level of noise (Subject 3 in Figure~\ref{invivo_predictions}). In contrast, our method appeared to be more robust to noise (as evident in Subjects 1, 3 and 4 in Figure~\ref{invivo_predictions}).

A critical concern with deep learning methods is their capacity to generalize. Models frequently fail to extrapolate when the distribution shift between training and validation datasets is too large. This issue is particularly pronounced in medical imaging, where distribution shifts can arise from various factors, such as differences in data acquisition or preprocessing, and training with unrealistic simulations. In our study, we obtained encouraging results regarding the generalization capability of our models, despite being trained exclusively on \textit{in silico} samples. Notably, our models demonstrated high-quality results on unseen non-flow \textit{in vitro} data and on \textit{in vivo} data. We only observed large errors in a limited number of cases (as exemplified by Subject 2 in Figure~\ref{invivo_predictions}).

This highlights the effectiveness of our simulation pipeline in generating color Doppler samples to build supervised training datasets when no ground truth is available. Our pipeline enables users to expand the range of clinical scenarios, including different echocardiographic views, transmission types (e.g. focused, diverging), and transducer types. It uses patient-derived Doppler echocardiographic data to model cardiac tissue and blood backscatterers. Although our focus was on 3-chamber views, other views could thus be explored. Additionally, the SIMUS simulator provides the flexibility to specify various probe and transmission parameters, allowing for the simulation of ultrasound signals with non-focused waves. This capability facilitates the generation of a vast pool of training data, enabling the creation of more generalized models. It is also worth noting that the favorable results obtained in non-flow data during the \textit{in vitro} disk experiments suggests the potential for directly applying our models to other modalities involving Doppler velocity estimation, such as tissue Doppler. 

The capacity for generalization is also contingent upon the size of the training dataset. Data augmentation is a crucial strategy for enhancing generalization ability, as it reduces the risk of overfitting the training data by increasing the size and diversity of the training dataset. While typical augmentation techniques, such as rotation or cropping, can be readily applied to images, they can also result in significant distortion of data such as I/Q signals. Therefore, we have developed tailored augmentation procedures to increase the size of the dataset while maintaining realistic signals. We observed experimentally that increasing the number of augmented samples was linked to enhanced generalization performances.

Finally, this work could offer practical advantages for focused color Doppler echocardiography. For instance, assuming a typical packet size of around $n=8$, reducing the packet size to $n=2$ could result in a fourfold increase in frame rate, yielding approximately $60$ frames/cardiac cycle. Such an enhancement could ease quantitative assessment of blood velocity throughout the cardiac cycle, potentially simplifying the application of techniques like intracardiac Vector Flow Mapping.

% network comparison
\subsection{Model comparison}
Two data representation methods were evaluated: the real-valued and complex-valued. Additionally, two types of architectures were benchmarked: U-Net and ConvNeXt.

\subsubsection{Real-valued vs. complex-valued representations}
We anticipated that Complex U-Net would outperform Real U-Net, in line with findings from our previous studies \cite{jingfeng}. However, Real U-Net and Complex U-Net yielded comparable results. One hypothesis is that while the Doppler velocity estimation problem involves complex-valued inputs, the output data are real-valued. Therefore, a transition from complex-valued to real-valued representations may be necessary, and a complex-valued architecture may not fully leverage its capabilities in this specific scenario. We also noted that the use of complex-valued representations significantly increased computational time during both training and inference due to the higher number of operations associated with complex-valued convolutions.

\subsubsection{U-Net vs ConvNeXt}
In this study, we analyzed ConvNeXt, a state-of-the-art deep learning architecture built on enhanced convolutional layers mirroring Transformer blocks for improved performance. We found similar performance from U-Net and ConvNeXt on \textit{in silico} and \textit{in vitro} data. Although ConvNeXt exhibited imperfect results on highly aliased \textit{in vitro} samples, it demonstrated greater robustness when applied to \textit{in vivo} aliased samples. This performance indicates the greater ability of ConvNeXt to learn intracardiac flow geometry from \textit{in silico} data. One of the reasons may be the larger number of training parameters of the ConvNeXt, allowed by the use of the efficient depthwise separable convolutions, which require fewer computational operations than standard convolutions. Hence, ConvNeXt may generalize better to intracardiac \textit{in vivo} data than to non-flow phantom data, which is highly out-of-distribution compared to training data. Moreover, despite including more parameters, ConvNeXt provided fast training and inference times.

% limitations and perspectives
\subsection{Limitations and perspectives}
Despite the positive findings presented in this paper, we acknowledge two main limitations of this work. First, the \textit{in vivo} results must be regarded as preliminary as the absence of ground truth made it difficult to draw firm conclusions on estimation quality. Second, we did not account for clutter influence and assumed that clutter filtering could be performed effectively on a packet size of two. A challenging perspective is to develop a clutter filtering method specifically designed for reduced packet sizes. We believe that exploring deep learning techniques could be a promising avenue for addressing this task.

Another perspective is to incorporate temporal information \cite{ling_segmentation}. At this stage, our simulations were based on patient data from an echocardiographic laboratory, and thus lacked sufficient temporal resolution to assume continuity between consecutive frames. Realistic color flow imaging simulations based on a numerical heart, such as those employed previously in our research \cite{sun, quarteroni_using_2015}, would provide the necessary frame rate to leverage temporal information effectively and improve the robustness of our models.

Lastly, our models are expected to generalize to pathological flows, especially if non-standard flows are present in the training dataset. However, due to anonymization, we cannot determine which or how many of the patients used to generate \textit{in silico} data have a cardiac condition. Also, the \textit{in vivo} validation was exclusively performed on healthy volunteers. Therefore, a validation study on a cohort including a variety of pathological cases could be envisaged for future work. In particular, it would allow us to assess the robustness of ConvNeXt when encountering highly abnormal flows, as it showed suboptimal performance when processing out-of-distribution \textit{in vitro} data.

%----------------------------------------------------------------------

\section{Conclusion}
\label{conclusion}
We explored deep learning models for Doppler velocity estimation from clutter-filtered I/Q signals with a reduced packet size of $n=2$. Using a color Doppler simulation pipeline, we built a supervised training dataset and applied custom data augmentation, including artificially adding aliased samples. Our method exhibited superior performance on \textit{in silico} data compared to the autocorrelation method, and achieved proficient results on \textit{in vitro} and \textit{in vivo} data. We demonstrated the effectiveness of our approach in reducing aliasing and handling noise. Three state-of-the-art deep learning strategies were adapted and evaluated, with ConvNeXt showing the best generalization performance on \textit{in vivo} samples. Overall, our study highlights the potential of deep learning for quantitative color Doppler analysis with minimal data.

%----------------------------------------------------------------------

\section*{Acknowledgments}
This work has benefited from access to IDRIS computing resources through the allocation of resources 2022-AD011013679 granted by GENCI. The \textit{in vitro} material is based upon work done on the ISO 9001:2015 PILoT facility. The RF Verasonics generator was cofounded by the FEDER program, Saint-Etienne Metropole (SME), and Conseil General de la Loire (CG42) within the framework of the SonoCardio-Protection Project leaded by Pr Pierre Croisille.

The authors would like to thank Adeline Bernard for the phantom design, Nin Ghigo for their help with \textit{in vivo} acquisitions, and Pr Olivier Bernard, Thierry Judge and Florian Vixège for their help in the development of the simulation pipeline.

% ---------------------------------------------------------

\bibliographystyle{IEEEtran}
\bibliography{tuffc}

\end{document}